\documentclass[aps,prb,twocolumn,showpacs,groupedaddress]{revtex4}

\usepackage{graphicx}

\begin{document}

\preprint{}

\title{Magnetic-field induced resistivity minimum with in-plane linear magnetoresistance of the Fermi liquid in SrTiO$_{3-\emph{x}}$ single crystals}

\author{Z. Q. Liu (Zhiqi Liu)$^{1,2}$}

\author{W. M. L\"{u}$^{1,3}$}

\author{X. Wang$^{1,2}$}

\author{Z. Huang$^{1}$}

\author{A. Annadi$^{1,2}$}

\author{S. W. Zeng$^{1,2}$}

\author{T. Venkatesan$^{1,2,3}$}

\author{Ariando$^{1,2}$}

\altaffiliation[Email: ]{ariando@nus.edu.sg}

\affiliation{$^1$NUSNNI-Nanocore, National University of Singapore, 117411 Singapore}

\affiliation{$^2$Department of Physics, National University of Singapore, 117542 Singapore}

\affiliation{$^3$Department of Electrical and Computer Engineering, National University of Singapore, 117576 Singapore}

\date{\today}

\begin{abstract}
We report novel magnetotransport properties of the low temperature
Fermi liquid in SrTiO$_{3-\emph{x}}$ single crystals. The classical
limit dominates the magnetotransport properties for a magnetic field
perpendicular to the sample surface and consequently a
magnetic-field induced resistivity minimum emerges. While for the
field applied in plane and normal to the current, the linear
magnetoresistance (MR) starting from small fields ($<$ 0.5 T)
appears. The large anisotropy in the transverse MRs reveals the
strong surface interlayer scattering due to the large gradient of
oxygen vacancy concentration from the surface to the interior of
SrTiO$_{3-\emph{x}}$ single crystals. Moreover, the linear MR in our
case was likely due to the inhomogeneity of oxygen vacancies and
oxygen vacancy clusters, which could provide experimental evidences
for the unusual quantum linear MR proposed by Abrikosov [A. A.
Abrikosov, Phys. Rev. B \textbf{58}, 2788 (1998)].
\end{abstract}



\maketitle

\section{Introduction}

SrTiO$_{3}$ (STO) is one of the most important workhorses in oxide
electronics. Recently, a two-dimensional electron gas [1,2] and
electronic phase separation [3] have been demonstrated to emerge on
the bare STO single crystal surface. Understanding the electronic
properties of STO under different oxidation states is therefore
crucial to reveal the origin of these novel phenomena and to use STO
in electronic devices. Generally, STO is a nonpolar band insulator
with an indirect bandgap of $\sim$3.27 eV [4] and a large dielectric
constant \emph{$\varepsilon$$_{r}$} [5]. A semiconducting (or
metallic) phase of STO can be obtained by reduction [6], chemical
doping [7] or photo-carrier injection [8], with a high carrier
mobility ($>$10,000 cm$^{2}$V$^{-1}$s$^{-1}$) at low temperatures, a
large density-of-states effective mass \emph{m$_{D}$} =
5$\sim$6\emph{m}$_{0}$ [9,10] and a large cyclotron mass
\emph{m$_{c}$} = 1.5$\sim$2.9\emph{m}$_{0}$ [10], where
\emph{m}$_{0}$ is the electron rest mass. The high mobility carriers
allow the observation of magnetic quantum effects like Shubnikov-de
Haas oscillation. However, the additional conditions [11]
\emph{E$_{F}$} =
($\hbar$$^{2}$/2\emph{m$_{D}$})(3$\pi$$^{2}$\emph{n})$^{2/3}$ $\gg$
\emph{kT} and $\hbar$\emph{w$_{c}$} = $\hbar$\emph{eB/m$_{c}$} $\gg$
\emph{kT} for a pronounced effect of quantization on
magnetotransport have also to be considered, where \emph{E$_{F}$} is
the Fermi energy, \emph{n} the carrier density, \emph{B} the applied
magnetic field and \emph{w$_{c}$} = \emph{eB/m$_{c}$} the cyclotron
frequency. Taking \emph{m$_{c}$} = 2\emph{m}$_{0}$ for example, the
magnetic energy $\hbar$\emph{w$_{c}$} at 9 T exceeds the thermal
energy \emph{kT} below $\sim$6 K.

\begin{figure}
\includegraphics[width=3.4in]{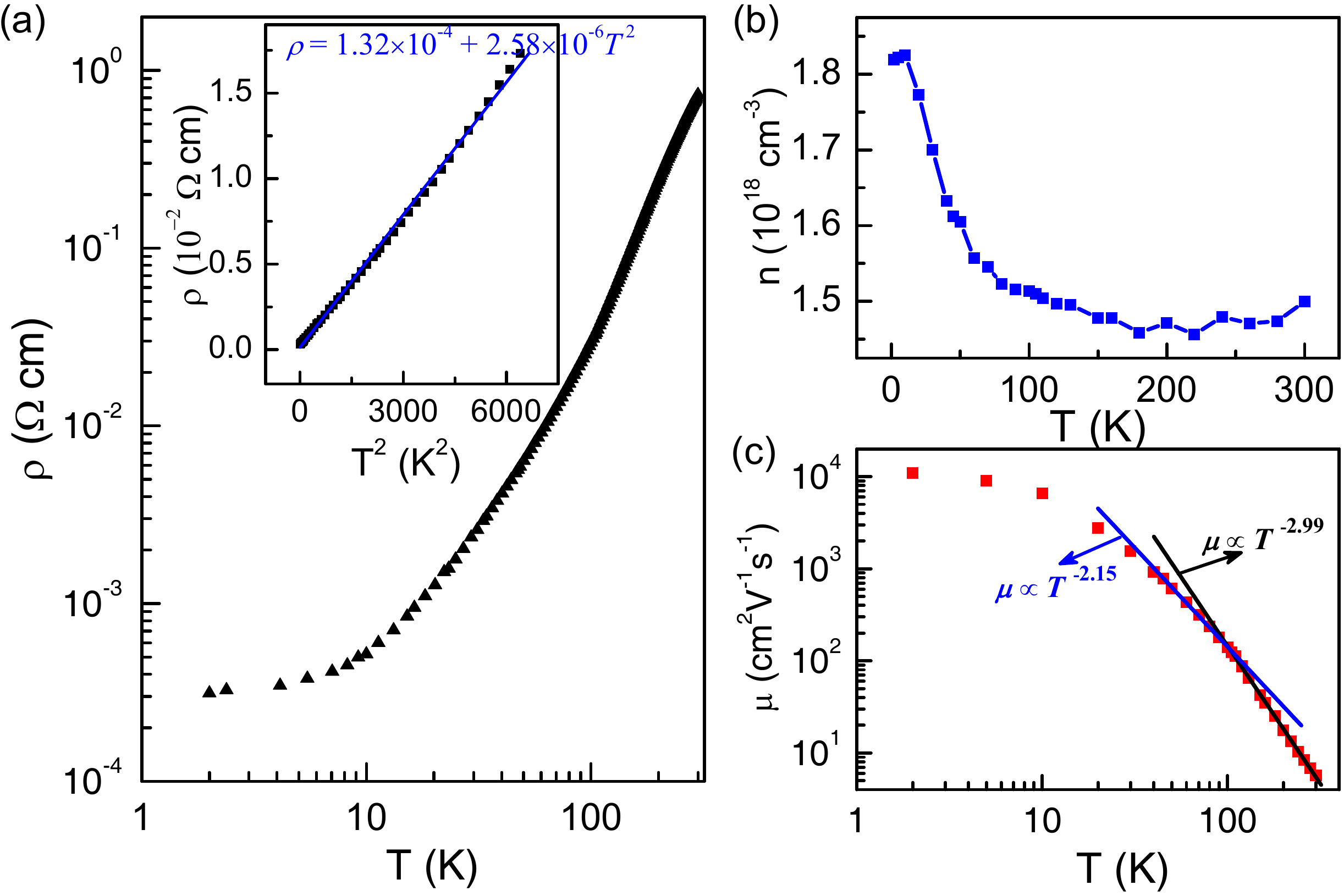}
\caption{\label{fig1} (Color online) Temperature dependence of (a)
resistivity (\emph{$\rho$-T}), (b) carrier density (\emph{n-T}) and
(c) mobility (\emph{$\mu$-T}) of a SrTiO$_{3}$ single crytal reduced
in $\sim$10$^{-7}$ Torr vacuum at 950 $^\circ$C for 1 h. Inset of
(a): linear fitting of \emph{T}$^{2}$ dependence of the resistivity.
The carrier density \emph{n} in (b) was averaged over the entire
crystal thickness.}
\end{figure}

In this paper we report electrical and magnetotransport studies of
SrTiO$_{3-\emph{x}}$ single crystals ($5\times5\times0.5$ mm$^3$)
which were reduced in $\sim$10$^{-7}$ Torr at 950$^\circ$C [6,12]
for different times. The electrical contacts were made by wire
bonding using aluminum wires. All the transport measurements were
performed in a Quantum Design Physical Property Measurement System.

\section{Electrical Properties}

The temperature dependence of resistivity (\emph{$\rho$-T}), carrier
density (\emph{n-T}) (\emph{n} is an average carrier density value
over the entire crystal thickness) and mobility (\emph{$\mu$-T}) of
samples reduced for 1 h is depicted in Fig.1, showing a metallic
behaviour over the whole temperature range from 300 to 2 K (Fig.
1(a)). The origin of this metallic behavior can be understood in
terms of the Mott criterion [13]. The critical carrier density for a
metal-insulator transition is given by the Mott critical carrier
density \emph{n$_{c}$} $\approx$ (0.25/\emph{a$^{\ast}$})$^{3}$,
where \emph{a$^{\ast}$} =
4$\pi$\emph{$\varepsilon$$_{r}$}\emph{$\varepsilon$}$_{0}$$\hbar$$^{2}$/\emph{m$_{D}$}\emph{e}$^{2}$
is the effective Bohr radius and \emph{$\varepsilon$}$_{0}$ the
vacuum permittivity. The measured carrier density at 300 K is
$\sim$1.5$\times$10$^{18}$ cm$^{-3}$, which is more than three times
\emph{n$_{c}$} $\sim$ 4.9$\times$10$^{17}$ cm$^{-3}$, considering
the room temperature \emph{$\varepsilon$$_{r}$} $\approx$ 300 and
\emph{m$_{D}$} $\approx$ 5\emph{m}$_{0}$ for STO. Interestingly, the
resistivity from 2 up to $\sim$80 K exhibits an obvious behavior of
a strongly correlated Fermi liquid as fitted in the inset of Fig.
1(a), reminiscent of the normal state of electron-doped cuprate
superconductors [14] and the \emph{p}-wave superconductor
Sr$_{2}$RuO$_{4}$ [15], noting that semiconducting STO is also
superconducting at very low temperatures [12]. Moreover, the common
Fermi liquid origin of the \emph{T}$^{2}$ resistivity and
superconductivity of \emph{n}-type SrTiO$_{3}$ has been elaborately
discussed by Marel \emph{et al}. [16].

The detailed \emph{n-T} curve determined by Hall measurements is
shown in Fig. 1(b). At high temperatures, the carrier density
slightly fluctuates; however, it increases with decreasing
temperature at low temperatures especially between 100 and 10 K.
This unexpected behavior is unphysical from the viewpoint of thermal
activation. In fact, this behavior was also observed previously in
semiconducting STO single crystals [7,8] and LaAlO$_{3}$/STO
heterostructures grown at low oxygen pressures [3,17]. Therefore
there should be another intrinsic mechanism affecting the carrier
density.

One unique property of STO is that its \emph{$\varepsilon$$_{r}$}
increases with lowering temperature (especially from $\sim$100 K)
and saturates at 4 K because of the quantum-mechanical stabilization
of the paraelectric phase [5]. It seems plausible to assume that in
SrTiO$_{3-\emph{x}}$ part of carriers are trapped by the Coulomb
potentials of the majority of positively charged defects due to the
strongly ionic nature of the lattice. As the
\emph{$\varepsilon$$_{r}$} increases, the Coulomb potentials will be
suppressed due to dielectric screening since the screened Coulomb
potential [18] is inversely proportional to
\emph{$\varepsilon$$_{r}$}. Hence the increase of
\emph{$\varepsilon$$_{r}$} could serve as a kind of detrapping
mechanism and consequently account for the increase of carrier
density in SrTiO$_{3-\emph{x}}$ at low temperatures.

The \emph{$\mu$-T} curve is plotted in Fig. 1(c) on a logarithmic
scale. The high mobility, up to $\sim$11,000
cm$^{2}$V$^{-1}$s$^{-1}$ at 2 K, decreases with temperature rapidly
and varies in accordance with certain power laws above 30 K, where
the scattering of electrons by polar optical phonons dominates and
results in $\mu$ $\propto$ \emph{T$^{-\beta}$} [19]. The effect of
the structural phase transition in STO at $\sim$105 K on the
mobility can be apparently seen from the linear fittings owing to
the variation in the activation of phonon modes. Below 30 K, the
scattering of electrons is dominated by ionized defect scattering
and electron-electron Umklapp scattering because of the Fermi liquid
behavior [15]. However, the electron-electron scattering should be
enhanced with increasing carrier density, so the ionized defect
scattering is the more pronounced mechanism for the continuously
increasing mobility similarly due to the dielectric screening of
ionized scattering potentials [7]. Finally, the trend to saturation
appears from 5 K possibly corresponding to the quantum paraelectric
phase in STO.

\begin{figure}
\includegraphics[width=3.4in]{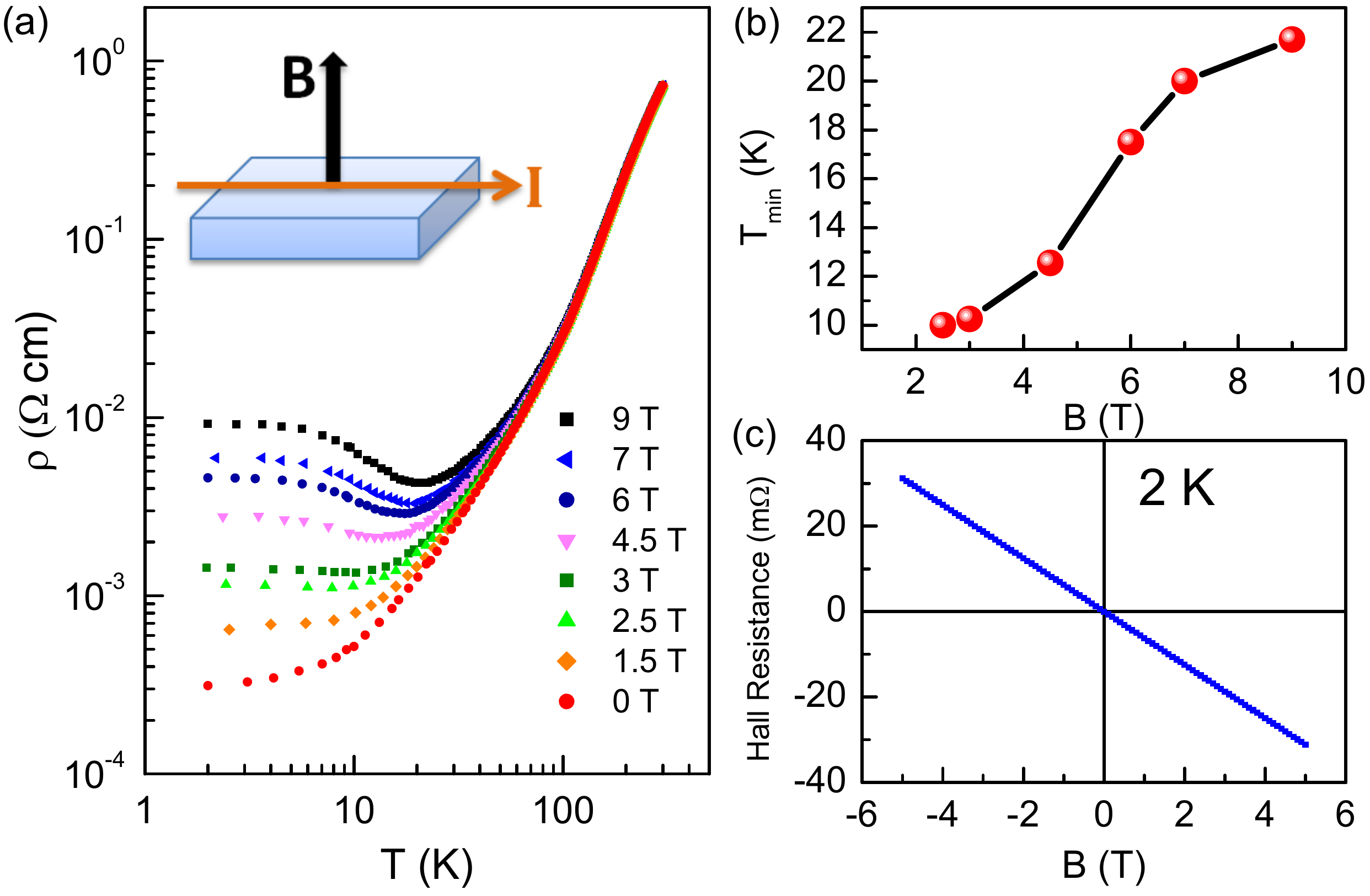}
\caption{\label{fig2} (Color online) (a) \emph{$\rho$-T} curves
under different magnetic fields. (b) Extracted resistivity minimum
temperature from (a) versus magnetic field. (c) Large field Hall
effect at 2 K from -5 to 5 T.}
\end{figure}

\section{Magnetotransport Properties}
The \emph{$\rho$-T} curves under different magnetic fields
perpendicular to the surface are shown in Fig. 2(a). The transverse
MR effect ($\triangle\rho$/$\rho$(0) = [$\rho$(B)-
$\rho$(0)]/$\rho$(0)) is notable only below $\sim$50 K and always
positive. Intriguingly, the \emph{$\rho$-T} curves under
sufficiently strong magnetic fields ($\geq$2.5 T) exhibit a
resistivity minimum at a temperature \emph{T$_{min}$}, which
increases with \emph{B} monotonically as plotted in Fig. 2(b). The
resistivity minimum cannot pertain to a Kondo effect or weak
localization since they are inherently antagonistic towards magnetic
fields. The antilocalization effect is also ruled out since it is
typically very small (of the order of few percents). One possible
origin for this behaviour could be magnetic field induced carrier
freeze-out due to the considerable shrinking of electron wave
functions if the magnetic field strength is much larger than the
Coulomb forces [20]. The magnetic freeze-out would cause an increase
of Hall coefficient. To examine this, Hall measurements was
performed up to 5 T at 2 K. However, the observed linear Hall effect
indicates that large magnetic fields are not affecting the carrier
density at all as seen in Fig. 2(c).

To explore the origin of the magnetic-field induced resistivity
minimum, the out-of-plane transverse MR was measured up to 9 T. As
shown in Fig. 3(a), both the 2 and 10 K MR curves exhibit an obvious
quadratic shape. For simplicity, we analyze our data using the
single band picture, where the quadratic MR can be described by the
classical orbital scattering. The Fermi energy at 2 K is
\emph{E$_{F}$} $\approx$ 1.08 meV and more than $\sim$6\emph{kT}(2
K) taking \emph{m$_{D}$} $\approx$ 5\emph{m}$_{0}$ and \emph{n} as
the average carrier density over the entire crystal thickness at 2 K
as shown in Fig. 1(b), corresponding to a degenerate gas state
(\emph{E$_{F}$} $\geq$ \emph{kT}). Indeed, the carrier density of
the surface region should be larger than the average value due to
the inhomogeneity of oxygen vacancies (this will be discussed
later), which would thus lead to an even larger \emph{E$_{F}$}.
However, no signature of quantum oscillations in MR was seen in
spite of the high mobility. This could be because both the
\emph{m$_{D}$} and \emph{m$_{c}$} have been largely enhanced from
strong electron correlations of the Fermi liquid and eventually the
initially assumed degenerate gas actually exists in a non-degenerate
(\emph{E$_{F}$} $\leq$ \emph{kT})``liquid" state. In addition, the
Shubnikov-de Haas oscillation was observed typically at lower
temperatures below 2 K [10].

The MR of 9 T at 2 K is extremely large, more than 2,900\% and
decreases to $\sim$1,200\% at 10 K. The classical transverse orbital
scattering can be described by $\triangle\rho$/$\rho$(0) =
$\alpha$$^{2}$$\mu$$^{2}$\emph{B}$^{2}$, where $\alpha$ is a
material dependent constant. By fitting the MR curves of 2 and 10 K,
the average value of $\alpha$ is obtained to be $\sim$0.32, which is
comparable to the $\alpha$ $\sim$ 0.38 in \emph{n}-type InSb [21].
The mobility of 2 K is larger than that of 10 K and therefore the MR
of 2 K is far larger as shown in Fig. 3(a). Under a sufficiently
large \emph{B}, the MR difference between 2 and 10 K is so large
that the overall resistivity
 $\rho$(\emph{B,T}) = $\rho$(0,\emph{T}) + $\alpha$$\mu$$^{2}$\emph{B}$^{2}$$\rho$(0,\emph{T}) at 2 K can become larger than that of
 10 K although $\rho$(0,2 K) is smaller than $\rho$(0,10 K) in the normal metallic state.
 In this way, the intriguing resistivity minimum under a large magnetic field can be understood.

\begin{figure}
\includegraphics[width=2.3in]{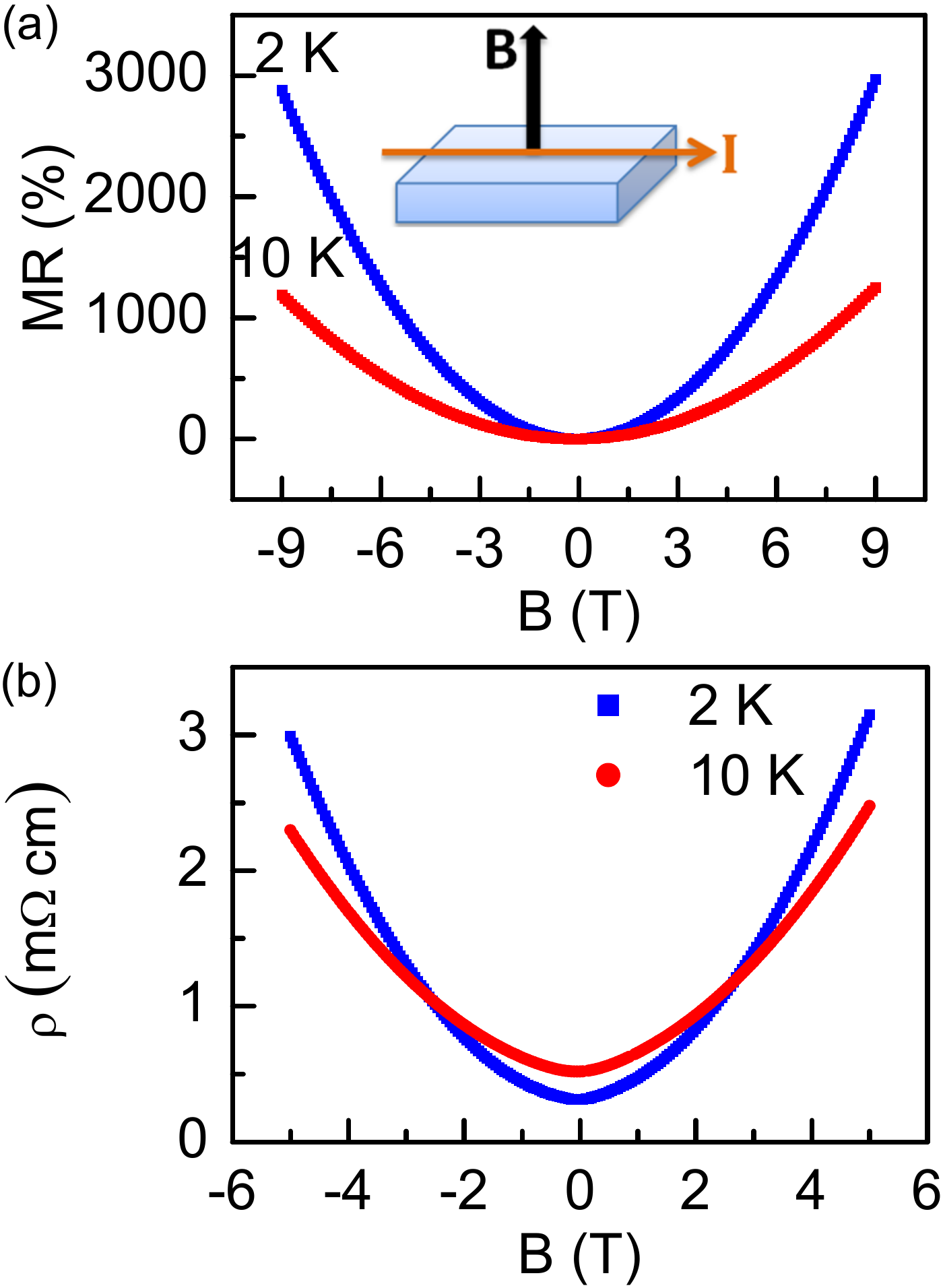}
\caption{\label{fig3}(Color online) (a) Out-of-plane
magnetoresistance (MR) at 2 and 10 K up to 9 T. Inset: schematic of
the measurement geometry. (b) Magnetic field dependence of the
resistivity (\emph{$\rho$-B}) for 2 and 10 K up to 5 T. }
\end{figure}

To further confirm the origin of this resistivity minimum, the
magnetic field dependence of resistivity (\emph{$\rho$-B}) at 2 and
10 K are compared in Fig. 3(b). There is an evident crossover
between the two \emph{$\rho$-B} curves at $\sim$2.5 T, above which
the resistivity at 2 K exceeds that at 10 K. The crossover field is
consistent with the critical \emph{B} in Fig. 2(a). The low
temperature resistivity $\rho$(\emph{B,T}) = $\rho$(0,\emph{T}) +
$\alpha$$\mu$$^{2}$\emph{B}$^{2}$$\rho$(0,\emph{T}) can be readily
simulated up to 80 K by considering the \emph{T}$^{2}$ dependence of
$\rho$(0,\emph{T}) and mathematically representing the mobility
below 30 K with an exponential fitting. The simulated results (not
shown) were consistent with the curves in Fig. 2(a), which suggests
that the above resistivity comparison is not only true between 2 and
10 K but also valid for other temperatures. Finally we conclude that
the magnetic field induced resistivity minimum originates from the
extremely large MR and its pronounced increase with decreasing
temperature at low temperatures. The large MR is achieved by the
fairly high $\mu$, sufficiently large \emph{B} and the possible
stabilization of the classical limit by strong electron correlations
of the Fermi liquid.

Similar behavior was also observed in SrTiO$_{3-\emph{x}}$ single
crystals reduced for 2 h, which possess a larger room temperature
carrier density and also a high mobility at 2 K. The Fermi liquid
behavior, \emph{i.e.} the \emph{T}$^{2}$ dependence of the
resistivity, also exists but up to $\sim$65 K. Nevertheless, as the
reducing time was prolonged to 8 h for a STO single crystal, the
room temperature carrier density reaches $\sim$6$\times$10$^{18}$
cm$^{-3}$ and consequently the mobility at 2 K is only $\sim$2,500
cm$^{2}$V$^{-1}$s$^{-1}$. As a result there is no observable
resistivity minimum even under a 9 T magnetic field.

The condition for the strong-field region $\hbar$\emph{w$_{c}$}
$\gg$ \emph{kT}, in which most of the carriers are in the lowest
Landau magnetic quantum level, was coined as the ``quantum limit"
[22]. The theoretical analyses [23-25] indicate that for both
degenerate and nondegenerate statistics the transverse MR has a
quadratic field dependence in the classical low-field case with
$\hbar$\emph{w$_{c}$} $\ll$ \emph{kT} but a linear dependence in the
quantum limit. Moreover, the other criterion \emph{n} $\ll$
(eB/$\hbar$)$^{3/2}$ [25] should also be fulfilled for the usual
quantum linear MR as observed in high mobility InSb [21,26], Ge
[27], graphite [28] and also recently in the topological insulator
Bi$_{2}$Se$_{3}$ [29]. As shown in Fig. 4(a), the MR curves exhibit
highly linear field dependence while the \emph{B} is applied in
plane and transverse to the current. The magnitude of the in-plane
MR is on average $\sim$92\% at 2 K and $\sim$66\% at 10 K under 9 T,
which are far smaller than the out-of-plane values. Accordingly
neither the crossover in the \emph{$\rho$-B} curves (inset of Fig.
4(a)) nor the magnetic field induced resistivity minimum was
observed. In our case, the linear MR starts from a very small field
$<$ 0.5 T as seen from Fig. 4(a), which in turn corresponds to a
critical carrier density of $\sim$1.36$\times$10$^{17}$ cm$^{-3}$
for the usual quantum linear MR. The average carrier density at 2 K
in our case is already $\sim$1.85$\times$10$^{18}$ cm$^{-3}$ and
more than one order of magnitude larger than the critical carrier
density. In this case, the linear MR may be out of the usual quantum
linear MR picture [30].

\begin{figure}
\includegraphics[width=2.3in]{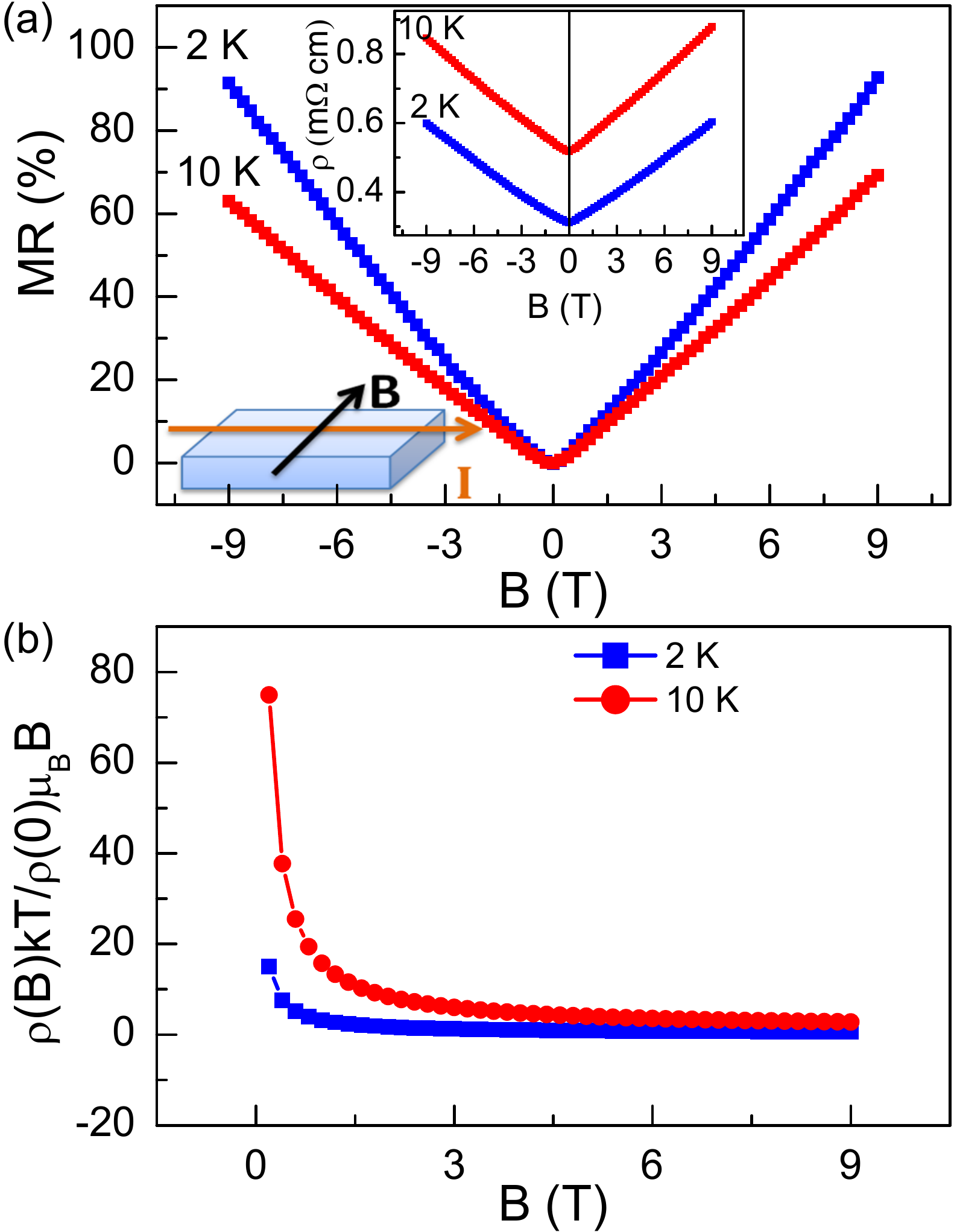}
\caption{\label{fig4} (Color online) (a) In-plane transverse MR at 2
K and 10 K up to 9 T. The upper and lower insets are the
corresponding \emph{$\rho$-B} curves of the two temperatures and the
schematic of measurement geometry respectively. (b) The parameter
$\frac{\rho(\emph{B})\cdot\emph{kT}}{\rho(0)\cdot\mu\emph{$_{B}$}\emph{B}}$
plotted as a function of magnetic field.}
\end{figure}

On the other hand, the linear MR starting from small fields was
previously also observed in nonstoichiometric silver chalcogenides
Ag$_{2+\emph{x}}$Se and Ag$_{2+\emph{x}}$Te by Xu \emph{et al.}
[31]. In that case, the criteria for the usual quantum linear MR can
also not be fulfilled. However, Abrikosov [25,32] proposed another
model of quantum linear MR for the Ag$_{2+\emph{x}}$Se and
Ag$_{2+\emph{x}}$Te case with two assumptions, \emph{i.e.}, (1) the
substance is inhomogenous, consisting of clusters of excess silver
atoms with a high electron density, surrounded by a medium with a
much lower electron density; (2) in this medium the electron energy
spectrum is close to a gapless semiconductor with a linear
dependence of energy on momentum, and thereby explained all the data
in [31] satisfactorily. Our scenario seems quite close to the above
case since the inhomogeneity of oxygen vacancies in
SrTiO$_{3-\emph{x}}$ single crystals has been a long-standing issue
[33,34]. Moreover, the oxygen vacancy clustering in
SrTiO$_{3-\emph{x}}$ has been well studied both theoretically by
Cuong \emph{et al}. [35] and experimentally by Muller \emph{et al.}
[36]. Similar to clusters of excess silver atoms in
nonstoichiometric silver chalcogenides, oxygen vacancy clusters have
a high electron density. Thus, the linear MR in our case could be
another example for the unusual quantum linear MR induced by
inhomogeneities and conductive clusters.

Considering the further spin-orbital splitting of the Landau
magnetic levels under a magnetic field \emph{B}, the energy scale
$\mu$\emph{$_{B}$}\emph{B} is involved, where $\mu$\emph{$_{B}$} is
the Bohr magneton. The dimensionless parameter
$\frac{\rho(\emph{B})\cdot\emph{kT}}{\rho(0)\cdot\mu\emph{$_{B}$}\emph{B}}$
is theoretically predicted to be relatively independent of
temperature for the quantum limit as illustrated in [26]. It is
plotted in Fig. 4(b) as a function of field for two different
temperatures and the parameter is approaching a constant value. This
further supports the idea that the linear in-plane transverse MR is
intrinsically a kind of quantum MR.

In early studies on the transport properties of semiconducting STO,
STO single crystals were thermally reduced in vacuum and high
temperature, which are similar to what we used, for quite a long
time (typically up to 10 days [6]) to achieve uniform oxygen vacancy
distribution. However, the oxygen vacancy distribution obtained by
that long time reduction was found [34] to be still not ideally
uniform over the entire thickness of single crystals. Furthermore,
the O$_{vac}$ in short-time (like 1 h in our case) reduced STO
single crystals are therefore far from uniform leading to a large
gradient in O$_{vac}$ concentration [34] and thus also in carrier
density [37] from the surface to the interior. Typically the oxygen
vacancy doping can increase carrier density but on the other hand
can decrease mobility in SrTiO$_{3-\emph{x}}$ due to
elctron-impurity scattering, enhanced electron-electron scattering
and suppressed screening effect. However, the lowering of mobility
by oxygen vacancies is not as significant as the increase of carrier
density by oxygen vacancy doping as can be seen from the electronic
transport property studies of SrTiO$_{3-\emph{x}}$ single crystals
by Frederikse \emph{et al.} [6] and SrTiO$_{3-\emph{x}}$ films by
Ohtomo and Hwang [38]. Thus, the higher oxygen vacancy concentration
results in smaller resistivity and accordingly the conducting
channel.

Consequently the measured transport data should only well represent
the properties of the surface layers for this kind of samples. The
highly anisotropic transport properties of the Fermi liquid can
therefore be understood by the strong surface interlayer scattering
due to the large inhomogeneity of O$_{vac}$ and accordingly the
density-of-states along the out-of-plane direction. Even though the
free carriers in our STO samples are not only confined to the very
top surface, the sharp gradient of the carriers with the largest
concentration at the top surface can result in the highly
anisotropic transport properties.

\section{Conclusions}
In conclusion, we studied the electrical and magnetotransport
properties of SrTiO$_{3-\emph{x}}$ single crystals. It was found
that the Fermi liquid exists at low temperatures and the dielectric
constant of STO plays an important role in carrier density and
mobility. A magnetic-field induced resistivity minimum was explored
to originate from the high mobility and the possible strengthening
of the classical limit by mass enhancement of strong electron
correlations. The linear in-plane transverse MR, potential for
linear MR sensors, was observed and attributed to the unusual
quantum linear MR due to the inhomogeneity of oxygen vacancies and
also oxygen vacancy clustering. The large anisotropy in transverse
MRs reveals the strong surface interlayer scattering due to the
inhomogeneity of O$_{vac}$ at the surface of SrTiO$_{3-\emph{x}}$.
By this work, we demonstrate a potential route to quantum linear MR
in virtue of inhomogeneities.

\begin{acknowledgments}
We thank the National Research Foundation (NRF) Singapore under the
Competitive Research Program (CRP) "Tailoring Oxide Electronics by
Atomic Control" NRF2008NRF-CRP002-024, National University of
Singapore (NUS) cross-faculty grant and FRC (ARF Grant No. R-144-000-278-112) for financial support.
\end{acknowledgments}


\end{document}